\documentclass{article}
\usepackage{amssymb, amsmath}

\input{tcilatex}
\begin{document}

\begin{titlepage}
\vspace*{3cm}
\begin{center}
{\Large \textsf{\textbf{Non-commutative space time of the relativistic equations with a Coulomb potential using Seiberg-Witten map}}}
\vskip 5mm
{\large \textsf{Slimane Zaim$^{*}$}}\\
\vskip 5mm
D\'{e}partement des Sciences de la Mati\`{e}re, Facult\'{e} des Sciences,\\
Universit\'{e} Hadj Lakhdar -- Batna, Algeria.\\
\vskip
2mm
{\large\textsf{\textbf{Abstract}}}
\end{center}
\begin{quote}
We
present an important contribution to the non-commutative approach to the
hydrogen atom
to deal with lamb shift corrections. This can be done by
studying the Klein-Gordon and Dirac equations
in a non-commutative
space-time up to first-order of the non-commutativity parameter using the
Seiberg-Witten maps.
We thus find the non-commutative modification of the
energy levels and by comparing with
the the current experimental results on
the Lamb shift of the 2P level
to extract a bound on the parameter of
non-commutativity, we show that
the fundamental length ($\sqrt{\Theta}$) is
compatible with the value of the electroweak length scale ($l$%
).
Phenomenologically, this effectively confirms the presence of gravity at
this level.
\end{quote}
\vspace*{2mm}
\noindent\textbf{\sc Keywords:}
non-commutative field theory,
Hydrogen atom, Klein-Gordon and Dirac
equations.

\noindent\textbf{\sc Pacs numbers}: 11.10.Nx, 32.30.-r ,
03.65.-w
\vskip 30mm
\noindent{\textsf{$^{*}$Corresponding Author, E-mail: 
\texttt{zaim69slimane@yahoo.com}}}
\end{titlepage}

\section{Introduction}

The standard concept of space-time as a geometric manifold is based on the
notion of a manifold whose points are locally labelled by a finite number of
real coordinates. However, it is generally believed that this picture of
space-time as a manifold should break down at very short distances of the
order of the Planck length. This implies that the mathematical concepts of
high energy physics has to be changed or more precisely our classical
geometric concepts may not be well-suited for the description of physical
phenomenon at short distances \cite{1,2,3}. The connection between string
theory and the non-commutativity \cite{4,5,6,7} motivated a large amount of
work to study and understand many physical phenomenon. The study of this
geometry has raised new physical consequences and thus, recently, a
non-commutative description of quantum mechanics has stimulated a large
amount of research \cite{8,9,10,11,12,13,14,15} . The non-commutative field
theory is characterised by the commutation relations between the position
coordinate operators themselves; namely: 
\begin{equation}
\left[ \hat{x}^{\mu },\hat{x}^{\nu }\right] =i\Theta ^{\mu \nu },
\end{equation}%
and the star Moyal product $\ast $ is defined between two fields $\psi
\left( x\right) $ and $\varphi \left( x\right) $ by: 
\begin{equation}
\psi \left( x\right) \ast \varphi \left( x\right) =\exp \left( \frac{i}{2}%
\Theta ^{\mu \nu }\frac{\partial }{\partial x^{\mu }}\frac{\partial }{%
\partial y^{\nu }}\right) \psi \left( x\right) \varphi \left( y\right) \mid
_{y=x},
\end{equation}%
where $\Theta ^{\mu \nu }$ are the non-commutativity constant parameters in
the canonical non-commutative space-time.

The issue of time-space non-commutativity is worth pursuing on its own right
because of its deep connection with such fundamental notions as unitarity
and causality. Much attention has been devoted in recent times to circumvent
these difficulties in formulating theories with $\Theta ^{0i}\neq 0$ \cite%
{1, 2, 16, 17}$.$ There are similar examples of theories with time-space
non-commutativity in the literature \cite{18, 19, 20} where unitarity is
preserved by a perturbative approach \cite{21}.

The most obvious natural phenomena to search for non-commutative effects are
simple systems of quantum mechanics in the presence a magnetic field, such
as a hydrogen atom. In the non-commutative time-space one expects the
degeneracy of the spectrum levels to be lifted, and therefore one can say
that the non-commutativity plays the role of the magnetic field. The study
of the exact and approximate solutions of relativistic hydrogen atom has
proved to be fruitful and many papers have been published \cite{22,23,24,25}%
. In this work we present an important contribution to the non-commutative
approach to the relativistic hydrogen atom. Our goal is to solve the
Klein-Gordan and Dirac equations for the Coulomb potential in a
non-commutative space-time up to first-order of the non-commutativity
parameter using the Seiberg-Witten maps and the Moyal product. We thus find
the non-commutative modification of the energy levels of the hydrogen atom
and we show that the non-commutativity is the source of a magnetic field
resulting in the lamb shift corrections. We also note that the effect of
non-commutativity confirms the presence of gravity at the very short
distances.

In a previous work \cite{26,27}, by solving the deformed Klein-Gordon and
Dirac equations in a canonical non-commutative space, we showed that the
energy is shifted, where the correction is proportional to the magnetic
quantum number, which behavior is similar to the Zeeman effect as applied to
a system without spin in a magnetic field, thus we explicitly accounted for
spin effects in this space.

The purpose of this paper is to study the extension of the Klein-Gordon and
Dirac fields in canonical non-commutative time-space by applying the result
obtained to a hydrogen atom.

This paper is organized as follows. In section $2$, we propose an invariant
action of the non-commutative boson and fermion fields in the presence of an
electromagnetic field. In section $3$, using the generalised Euler-Lagrange
field equations, we derive the deformed Klein-Gordon (KG) and Dirac
equations for the hydrogen atom. We solve these deformed equations and
obtain the non-commutative modification of the energy levels. Furthermore,
we derive the non-relativistic limit of the non-commutative KG equation for
a hydrogen atom and solve it using perturbation theory. Finally, in section $%
4$, we draw our conclusions.

\section{Action}

\bigskip The canonical non-commutative space-time is characterized by the
commutation relations of coordinate operators satisfying the relation ($1$).
In order to preserve this relation, the infinitesimal gauge transformation
is generalized by the following relation: 
\begin{equation}
\hat{\phi}^{A}\left( A\right) +\hat{\delta}_{\hat{\lambda}}\hat{\phi}%
^{A}\left( A\right) =\hat{\phi}^{A}\left( A+\delta _{\lambda }A\right) ,
\label{eq:trans}
\end{equation}
where $\hat{\phi}^{A}=(\hat{A}_{\mu },\hat{\psi})$ is a non-commutative
generic field, $\hat{A}_{\mu }$ and $\hat{\psi}$ are the non-commutative
gauge and matter fields respectively, $\lambda $ is the $\mathrm{U} (1)$
gauge Lie-valued infinitesimal transformation parameter, $\delta _{\lambda }$
is the ordinary gauge transformation and $\hat{\delta}_{\hat{ \lambda}}$ is
a non-commutative gauge transformation which are defined by: 
\begin{align}
\hat{\delta}_{\hat{\lambda}}\hat{\psi}=i\hat{\lambda}\ast \hat{\psi},&
\qquad \delta _{\lambda }\psi =i\lambda \psi ,  \label{eq:tempo1} \\
\hat{\delta}_{\hat{\lambda}}\hat{A}_{\mu }=\partial _{\mu }\hat{\lambda}+i%
\left[ \hat{\lambda},\hat{A}_{\mu }\right] _{\ast },& \qquad \delta
_{\lambda }A_{\mu }=\partial _{\mu }\lambda .  \label{tempo2}
\end{align}

Now using these transformations one can get at second order in the
non-commutative parameter $\Theta ^{\mu \nu }$ the following Seiberg--Witten
maps \cite{4}: 
\begin{align}
\hat{\psi}& =\psi +\psi ^{1}+\mathcal{O}\left( \Theta ^{2}\right) , \\
\hat{\lambda}& =\lambda +\lambda ^{1}\left( \lambda ,A_{\mu }\right) +%
\mathcal{O}\left( \Theta ^{2}\right) , \\
\hat{A}_{\xi }& =A_{\xi }+A_{\xi }^{1}\left( A_{\xi }\right) +\mathcal{O}%
\left( \Theta ^{2}\right) ,  \label{eq:SWM1} \\
\hat{F}_{\mu \xi }& =F_{\mu \xi }\left( A_{\xi }\right) +F_{\mu \xi
}^{1}\left( A_{\xi }\right) +\mathcal{O}\left( \Theta ^{2}\right) ,
\label{eq:SWM2}
\end{align}%
where: 
\begin{eqnarray}
\psi ^{1} &=&-\frac{i}{2}\Theta ^{\alpha \beta }(\left\{ A_{\alpha
},\partial _{\beta }\psi \right\} +\frac{1}{2}\left\{ \left[ \psi ,A_{\alpha
}\right] ,A_{\beta }\right\} ), \\
\lambda ^{1} &=&\Theta ^{\alpha \beta }\partial _{\alpha }\lambda A_{\beta },
\\
A_{\xi }^{1} &=&\frac{1}{2}\Theta ^{\alpha \beta }A_{\alpha }\left( \partial
_{\xi }A_{\beta }-2\partial _{\beta }A_{_{\xi }}\right) , \\
F_{\mu \xi }^{1} &=&-\Theta ^{\alpha \beta }\left( A_{\alpha }\partial
_{\beta }F_{\mu \xi }+F_{\mu \alpha }F_{\beta \xi }\right) ,
\end{eqnarray}%
and 
\begin{equation}
F_{\mu \nu }=\partial _{\mu }A_{\nu }-\partial _{\nu }A_{\mu }.
\end{equation}

To begin, we consider an action for a free boson and fermion fields in the
presence of an electrodynamic gauge field in a non-commutative space-time.
We propose the following action \cite{28a}: 
\begin{equation}
S=\int d^{4}x\,\left( \mathcal{L}_{\mathrm{MB}}+\mathcal{L}_{\mathrm{MF}}-%
\frac{1}{4}\hat{F}_{\mu \nu }\ast \hat{F}^{\mu \nu }\right) ,
\end{equation}%
where $\mathcal{L}_{\mathrm{MB}}$ and $\mathcal{L}_{\mathrm{MF}}$ are the
boson and fermion matter densities respectively in the non-commutative
space-time and are given by: 
\begin{equation}
\mathcal{L}_{\mathrm{MB}}=\eta ^{\mu \nu }\left( \hat{D}_{\mu }\hat{\varphi}%
\right) ^{\dagger }\ast \hat{D}_{\nu }\hat{\varphi}+m^{2}\hat{\varphi}%
^{\dagger }\ast \hat{\varphi},
\end{equation}%
and: 
\begin{equation}
\mathcal{L}_{\mathrm{MF}}=\overline{\hat{\psi}}\ast \left( i\gamma ^{\nu }%
\hat{D}_{\nu }-m\right) \ast \hat{\psi},
\end{equation}%
where the gauge covariant derivative is defined as: $\hat{D}_{\mu }=\partial
_{\mu }+ie\hat{A}_{\mu }$.

From the action variational principle the generalised equations of Lagrange
up to $\mathcal{O}\left( \Theta ^{2}\right) $ are \cite{28}: 
\begin{equation}
\frac{\partial \mathcal{L}}{\partial \hat{\Phi}}-\partial _{\mu }\frac{%
\partial \mathcal{L}}{\partial \left( \partial _{\mu }\hat{\Phi}\right) }%
+\partial _{\mu }\partial _{\nu }\frac{\partial \mathcal{L}}{\partial \left(
\partial _{\mu }\partial _{\nu }\hat{\Phi}\right) }+\mathcal{O}\left( \Theta
^{2}\right) =0,  \label{eq:field}
\end{equation}%
where: 
\begin{equation}
\mathcal{L=L}_{\mathrm{MB}}+\mathcal{L}_{\mathrm{MF}}-\frac{1}{4}\hat{F}%
_{\mu \nu }\ast \hat{F}^{\mu \nu }.
\end{equation}

\section{Non-commutative time-space KG equation}

Using the modified field equation \eqref{eq:field} , with the generic boson
field $\hat{\varphi}$ one can find in a free non-commutative space-time and
in the presence of the external potential $\hat{A}_{\mu }$ the following
modified Klein-Gordon equation: 
\begin{equation}
\left( \eta ^{\mu \nu }\partial _{\mu }\partial _{\nu }-m_{e}^{2}\right) 
\hat{\varphi}\,+\left( ie\eta ^{\mu \nu }\partial _{\mu }\hat{A}_{\nu
}-e^{2}\eta ^{\mu \nu }\hat{A}_{\mu }\ast \hat{A}_{\nu }+2ie\eta ^{\mu \nu }%
\hat{A}_{\mu }\partial _{\nu }\right) \hat{\varphi}=0,
\end{equation}%
where the deformed external potential $\hat{A}_{\mu }$ $\binom{-e/r}{0}$ in
free non-commutative space-time is \cite{29}: 
\begin{align}
\hat{a}_{0}& =-\frac{e}{r}-\frac{e^{3}}{\,r^{4}}\Theta ^{0k}x_{k}+\mathcal{O}%
\left( \Theta ^{2}\right) , \\
\hat{a}_{i}& =\frac{e^{3}}{4\,r^{4}}\Theta ^{ik}x_{k}+\mathcal{O}\left(
\Theta ^{2}\right) ,
\end{align}%
for a non-commutative time-space, $\Theta ^{0k}\neq 0$ and $\Theta ^{ki}=0$,
where $i,k=1,2,3$. In this case, we can check that: 
\begin{equation}
\eta ^{\mu \nu }\partial _{\mu }\partial _{\nu }=-\partial _{0}^{2}+\Delta ,
\end{equation}%
and 
\begin{equation}
2ie\eta ^{\mu \nu }\hat{A}_{\mu }\partial _{\nu }=i\frac{2e^{2}}{r}\partial
_{0}+2i\frac{e^{4}}{\,r^{4}}\Theta ^{0j}x_{j}\partial _{0},
\end{equation}%
and 
\begin{equation}
-e^{2}\eta ^{\mu \nu }\hat{A}_{\mu }\ast \hat{A}_{\nu }=\frac{e^{4}}{r^{2}}+2%
\frac{e^{6}}{\,r^{5}}\Theta ^{0j}x_{j},
\end{equation}%
then the Klein-Gordon equation $(20)$ up to $\mathcal{O}\left( \Theta
^{2}\right) $ takes the form: 
\begin{equation}
\left[ -\partial _{0}^{2}+\Delta -m_{e}^{2}+\frac{e^{4}}{r^{2}}+i\frac{2e^{2}%
}{r}\partial _{0}+2i\frac{e^{4}}{\,r^{4}}\Theta ^{0j}x_{j}\partial _{0}+2%
\frac{e^{6}}{\,r^{5}}\Theta ^{0j}x_{j}\right] \hat{\varphi}=0.
\end{equation}%
The solution to eq. $(26)$ in spherical polar coordinates $(r,\theta ,\phi )$
takes the separable form: 
\begin{equation}
\hat{\varphi}(r,\theta ,\phi ,t)=\frac{1}{r}\hat{R}(r)\hat{Y}\left( \theta
,\phi \right) \exp (-iEt).
\end{equation}%
Then eq. $(26)$ reduces to the radial equation: 
\begin{multline}
\left[ \frac{d^{2}}{dr^{2}}-\frac{l(l+1)-e^{4}}{r^{2}}+\frac{2Ee^{2}}{r}%
+\right. \\
\left. +E^{2}-m_{e}^{2}+2E\frac{e^{4}}{\,r^{4}}\Theta ^{0j}x_{j}+2\frac{e^{6}%
}{\,r^{5}}\Theta ^{0j}x_{j}\right] \hat{R}(r)=0.
\end{multline}

In eq.$(28)$ the coulomb potential in non-commutative space-time appears
within the perturbation terms \cite{30}: 
\begin{equation}
H_{\text{pert}}^{\Theta }=2E\frac{e^{4}}{\,r^{4}}\Theta ^{0j}x_{j}+2\frac{%
e^{6}}{\,r^{5}}\Theta ^{0j}x_{j},  \label{eq:perturbation}
\end{equation}%
where the first term is the electric dipole--dipole interaction created by
the non-commutativity, the second term is the electric dipole--quadruple
interaction. These interactions show us that the effect of space-time
non-commutativity on the interaction of the electron and the proton is
equivalent to an extension of two nuclei interactions at a considerable
distance. This idea effectively confirms the presence of gravity at this
level. To investigate the modification of the energy levels by eq. %
\eqref{eq:perturbation}, we use the first-order perturbation theory. The
spectrum of $H_{0}$ and the corresponding wave functions are well-known and
given by: 
\begin{equation}
R_{nl}(r)=\sqrt{\frac{a}{n+\nu +1}}\left( \frac{n!}{\Gamma \left( n+2\nu
+2\right) }\right) ^{1/2}x^{\nu +1}e^{-x/2}L_{n}^{2\nu +1}(x)\,,
\end{equation}%
where the relativistic energy levels are given by: 
\begin{equation}
E=E_{n,l}=\frac{m_{e}\left( n+\frac{1}{2}+\sqrt{\left( l+\frac{1}{2}\right)
^{2}-\alpha ^{2}}\right) }{\left[ \left( n+\frac{1}{2}\right) ^{2}+\left( l+%
\frac{1}{2}\right) ^{2}+2\left( n+\frac{1}{2}\right) \sqrt{\left( l+\frac{1}{%
2}\right) ^{2}-\alpha ^{2}}\right] ^{\frac{1}{2}}}\,,
\end{equation}%
and $L_{n}^{2\nu +1}$ are the associated Laguerre polynomials \cite{31},
with the following notations: 
\begin{equation}
\nu =-\frac{1}{2}+\sqrt{\left( l+\frac{1}{2}\right) ^{2}-\alpha ^{2}},\qquad
\alpha =e^{2},\qquad a=\sqrt{m_{e}^{2}-E^{2}}.
\end{equation}

\subsection{Non-commutative corrections of the relativistic energy}

Now to obtain the modification to the energy levels as a result of the terms %
\eqref{eq:perturbation} due to the non-commutativity of space-time we use
perturbation theory. For simplicity, first of all, we choose the coordinate
system $\left( t,r,\theta ,\varphi \right) $ so that $\Theta ^{0j}=-\Theta
^{j0}=\Theta \delta ^{01},$ such that $\Theta ^{0j}x_{j}=\Theta r$ and
assume that the other components are all zero and also the fact that in
first-order perturbation theory the expectation value of $1/r^{3}$ and $%
1/r^{4}$ are as follows: 
\begin{align}
\langle nlm&\mid r^{-3}\mid nlm^{\prime }\rangle =\int_{0}^{\infty
}R_{nl}^{2}(r)r^{-3}dr\delta _{mm^{\prime }}  \notag \\
&=\frac{4a^{3}n!}{\left( n+\nu +1\right) \Gamma \left( n+2\nu +2\right) }
\int_{0}^{\infty }x^{2\nu -1}e^{-x}\left[ L_{n}^{2\nu +1}(x)\right]
^{2}dx\delta _{mm^{\prime }}  \notag \\
&=\frac{4a^{3}n!}{\left( n+\nu +1\right) \Gamma \left( n+2\nu +2\right) } %
\left[ \frac{\Gamma \left( n+2\nu +2\right) }{\Gamma \left( n+1\right)
\Gamma \left( 2\nu +2\right) }\right] ^{2}\times  \notag \\
&\qquad\times \int_{0}^{\infty }x^{2\nu -1}e^{-x} \left[ F(-n;2\nu +2;x)%
\right] ^{2}dx\delta _{mm^{\prime }}  \notag \\
&=\frac{2a^{3}}{\nu \left( 2\nu +1\right) \left( n+\nu +1\right) }\left\{ 1+ 
\frac{n}{\left( \nu +1\right) }\right\} \delta _{mm^{\prime }}=f(3), \\
\langle nlm & \mid r^{-4}\mid nlm^{\prime }\rangle =\frac{4a^{4}}{\left(
2\nu -1\right) \nu \left( 2\nu +1\right) \left( n+\nu +1\right) }\left[ 1+%
\frac{3n }{\left( \nu +1\right) }\right. +  \notag \\
&\left. +\frac{3n\left( n-1\right) }{\left( \nu +1\right) \left( 2\nu
+3\right) }\right] \delta _{mm^{\prime }} =f(4),
\end{align}

Now, the correction to the energy to first order in $\Theta $ is: 
\begin{equation}
E^{\Theta \left( 1\right) }=\langle \psi _{nlm}^{0}\left\vert H_{\text{pert}
}^{\Theta \left( 1\right) }\right\vert \psi _{nlm}^{0}\rangle .
\end{equation}
where $H_{\text{pert}}^{\Theta \left( 1\right) }$ is the non-commutative
correction to the first order in $\Theta $ of the perturbation Hamiltonian,
which is given in the following relation: 
\begin{equation}
H_{\text{pert}}^{\Theta \left( 1\right) }=2E\frac{e^{4}}{\,r^{3}}\Theta +2%
\frac{e^{6}}{\,r^{4}}\Theta
\end{equation}

To calculate $E^{\Theta \left( 1\right) }$, we use the radial function in
equation $(30)$, to obtain: 
\begin{equation*}
E^{\Theta \left( 1\right) }=2\Theta \alpha ^{2}\left( E_{n,l}^{0}f\left(
3\right) +\alpha f\left( 4\right) \right)
\end{equation*}
Finally, the energy correction of the hydrogen atom in the framework of the
non-commutative KG equation is: 
\begin{eqnarray}
\Delta E^{\mathrm{NC}} &=&\frac{E^{\Theta \left( 1\right) }}{2E}  \notag \\
&=&\Theta \alpha ^{2}\left( f\left( 3\right) +\frac{\alpha }{E_{n,l}^{0}}%
f\left( 4\right) \right)
\end{eqnarray}

This result is important because it reflects the existence of Lamb shift,
which is induced by the non-commutativity of the space. Obviously, when $%
\Theta =0,$ then $\Delta E^{\mathrm{NC}}=0$, which is exactly the result of
the space-space commuting case, where the energy-levels are not shifted.

We showed that the energy-level shift for $1S$ is: 
\begin{equation}
\Delta E_{1S}^{\mathrm{NC}}=\Theta \alpha ^{2}\left( f_{1S}\left( 3\right) +%
\frac{\alpha }{E_{1,0}^{0}}f_{1S}\left( 4\right) \right)
\end{equation}
In out analysis, we simply identify spin up if the non-commutativity
parameter takes the eigenvalue $+\Theta $ and spin down if non-commutativity
parameter takes the eigenvalue $-\Theta $. Also we can say that the Lamb
shift is actually induced by the space-time non-commutativity which plays
the role of a magnetic field and spin in the same moment (Zemann effect).
This represents Lamb shift corrections for $l=0$. This result is very
important: as a possible means of introducing electron spin we replace $%
l\rightarrow \pm \left( j+\frac{1}{2}\right) $ and $n\rightarrow n-j-1-\frac{%
1}{2}$, where $j$ is the quantum number associated to the total angular
momentum. Then the $l=0$ state has the same total quantum number $j=\frac{1}{%
2}$. In this case the non-commutative value of the energy levels indicates
the splitting of $1s$ states.

\subsection{Non-relativistic limit}

The non-relativistic limit of the non-commutative K-G equation $(26)$ is
written as \cite{32,33}: 
\begin{equation}
\left[ \frac{d^{2}}{dr^{2}}-\frac{l(l+1)}{r^{2}}+\frac{2m_{e}e^{2}}{r}%
+2m_{e}\epsilon +2m_{e}\frac{e^{4}}{\,r^{3}}\Theta +2\frac{e^{6}}{\,r^{4}}%
\Theta \right] \hat{R}(r)=0.
\end{equation}%
In this non-relativistic limit the charged boson does not represent a single
charged particle, but is a distribution of positive and negative charges
which are different and extended in space linearly in $\sqrt{\Theta }$. The
absence of a perturbation term of form $\Theta /r^{2}$ in the
non-commutative coulomb interaction demonstrates that the distribution of
positive and negative charges is spherically symmetric. This can be
interpreted as the spherically symmetric distribution of charges of the
quarks inside in the proton.

Now to obtain the modification of energy levels as a result of the
non-commutative terms in eq. $(39)$, we use the first-order perturbation
theory. The spectrum of $H_{0}\left( \Theta =0\right) $ and the
corresponding wave functions are well-known and given by: 
\begin{equation}
\varepsilon _{n}=-\frac{m_{e}\alpha ^{2}}{2\hbar ^{2}n^{2}},
\end{equation}%
and 
\begin{equation}
R_{nl}(r)=\frac{1}{n}\left( \frac{\left( n-l-1\right) !}{a\left( n+l\right) !%
}\right) ^{1/2}x^{l+1}e^{-x/2}L_{n-l-1}^{2l+1}(x),\qquad x=\frac{2}{an}r,
\label{eq:solution0}
\end{equation}%
where $a=\hbar ^{2}/(m_{e}\alpha )$ is the Bohr radius of the Hydrogen atom.
The coulomb potential in non-commutative space-time appears within the
perturbation terms: 
\begin{equation}
H_{\text{pert}}^{\Theta }=2\Theta \alpha ^{2}\left( \frac{m_{e}}{\,r^{3}}+%
\frac{\alpha }{\,r^{4}}\right) +\mathcal{O}\left( \Theta ^{2}\right) ,
\end{equation}%
where the expectation values of $1/r^{3}$ and $1/r^{4}$ are as follows: 
\begin{eqnarray}
\langle nlm\mid r^{-3}\mid nlm^{\prime }\rangle _{l>0} &=&\frac{2}{%
a^{3}n^{3}l(l+1)(2l+1)}\delta _{mm^{\prime }}, \\
\langle nlm\mid r^{-4}\mid nlm^{\prime }\rangle _{l>0} &=&\left[ \frac{%
4\left( 3n^{2}-l(l+1)\right) }{a^{4}n^{5}l(l+1)(2l-1)(2l+1)(2l+3)}\right. 
\notag \\
&&\quad \left. +\frac{35\left( 3n^{2}-l(l+1)\right) }{%
3(l-1)(l+2)(2l-1)(2l+1)(2l+3)}\right] \delta _{mm^{\prime }}.
\end{eqnarray}%
Hence the modification to the energy levels is given by: 
\begin{equation}
\Delta E^{\mathrm{NC}}=\Theta \alpha ^{2}\left[ f\left( 3\right) +\frac{%
\alpha }{m_{e}}f\left( 4\right) \right] +\mathcal{O}\left( \Theta
^{2}\right) .
\end{equation}%
We can also compute the correction to the Lamb shift of the $2P$ level where
we have: 
\begin{equation}
\Delta E_{2P}^{\mathrm{NC}}=0.243\,156\,\Theta \left( \mathrm{MeV}\right)
^{3}.
\end{equation}

According to ref. \cite{34} the current theoretical result for the lamb
shift is $0.08\,\mathrm{kHz}$. From the splitting $(46)$, this then gives
the following bound on $\Theta $: 
\begin{equation}
\Theta \leq \left( 8.5\,\mathrm{TeV}\right) ^{-2}.
\end{equation}%
This corresponds to a lower bound for the energy scale of $8.5$ TeV, which
is in the range that has been obtained in refs \cite{35,36,37,38}, namely $%
1-10$ TeV.

\section{ Non-commutative time-space Dirac equation}

Now, concerning the Dirac equation in the free non-commutative time-space
and in the presence of the vector potential $\hat{A}_{\mu }$ and using the
modified field equation \eqref{eq:field}, with the generic field $\hat{\psi}$
we can find the modified Dirac equation up to $\mathcal{O}\left( \Theta
^{2}\right) $ as: 
\begin{equation}
\left( i\gamma ^{\mu }\partial _{\mu }-m_{e}\right) \hat{\psi}-e\gamma ^{\mu
}A_{\mu }\hat{\psi}-e\gamma ^{\mu }A_{\mu }^{1}\hat{\psi}+\frac{ie}{2}\Theta
^{\alpha \beta }\gamma ^{\mu }\partial _{\alpha }A_{\mu }\partial _{\beta }%
\hat{\psi}=0.  \label{eq:KGmod}
\end{equation}
For a non-commutative time-space ($\Theta ^{ki}=0$, where $i,k=1,2,3$), in
this case we can write: 
\begin{align}
i\gamma ^{\mu }\partial _{\mu }-m_{e}& =i\gamma ^{0}\partial _{0}+i\gamma
^{i}\partial _{i}-m_e, \\
-e\gamma ^{\mu }\hat{A}_{\mu }& =\frac{e^{2}}{r}\gamma ^{0}+\frac{e^{4}}{%
r^{4}}\gamma ^{0}\Theta ^{0k}x_{k}, \\
\frac{ie}{2}\Theta ^{\alpha \beta }\gamma ^{\mu }\partial _{\alpha }A_{\mu
}\partial _{\beta }& =-i\frac{e^{2}}{2}\gamma ^{0}\frac{\Theta ^{0k}x_{k}}{%
r^{3}}\partial _{0}.
\end{align}
Then the non-commutative Dirac equation \eqref{eq:KGmod} up to $\mathcal{O}
\left( \Theta ^{2}\right) $ takes the following form: 
\begin{equation}
\left[ i\gamma ^{0}\partial _{0}+i\gamma ^{i}\partial _{i}-m_{e}+\frac{e^{2} 
}{r}\gamma ^{0}+\frac{e^{4}}{r^{4}}\gamma ^{0}\Theta ^{0k}x_{k}-i\frac{e^{2} 
}{2}\gamma ^{0}\frac{\Theta ^{0k}x_{k}}{r^{3}}\partial _{0}\right] \hat{ \psi%
}=0.
\end{equation}
We can write this equation as: 
\begin{equation}
\hat{H}\hat{\psi}\left( t,r,\theta ,\varphi \right) =i\partial _{0}\hat{\psi}%
\left( t,r,\theta ,\varphi \right) .  \label{aa}
\end{equation}
Then replacing: 
\begin{equation}
\hat{\psi}\left( t,r,\theta ,\varphi \right)\to \exp \left( -iEt\right) \hat{
\psi}\left( r,\theta ,\varphi \right),
\end{equation}%
gives the stationary non-commutative Dirac equation: 
\begin{equation*}
\hat{H}\hat{\psi}\left(r,\theta ,\varphi \right) =E\hat{\psi}\left( r,\theta
,\varphi \right) ,
\end{equation*}
where $E$ is the ordinary energy of the electron and $\hat{H}$ is the
non-commutative Hamiltonian of the form: 
\begin{equation}
\hat{H}=\hat{H}_{0}+\hat{H}_{\mathrm{pert}}^{\Theta },
\end{equation}
where $H_{0}$ is the relativistic hydrogen atom Hamiltonian: 
\begin{equation}
\hat{H}_{0}=\vec{\boldsymbol{\alpha}}.\left(-i\vec{\boldsymbol{\nabla }}%
\right) +\beta m_{e}-\frac{e^{2}}{r},  \label{a1}
\end{equation}%
and $H_{\mathrm{pert}}^{\Theta }$ is the leading-order perturbation: 
\begin{equation}
\hat{H}_{\mathrm{pert}}^{\Theta }=\left( \frac{E}{2}-\frac{e^{2}}{r}\right)
e^{2}\frac{ {\vec{\boldsymbol{\Theta}} _{t}}\cdot \vec{\boldsymbol{r}}}{r^{3}%
}.  \label{a2}
\end{equation}

The leading long-distance part of $H_{\mathrm{pert}}^{\Theta }$ behaves like
that of a magnetic dipole potential where the non-commutativity plays the
role of a magnetic moment. So the non-commutative coulomb potential is the
multipolar contribution and this means that the distribution is not
spherically symmetric. In the above the matrices $\vec{\boldsymbol{\alpha }}$
and $\beta $ are given by: 
\begin{equation*}
\beta =\left( 
\begin{array}{cc}
I & 0 \\ 
0 & -I%
\end{array}
\right) ,\qquad \alpha ^{i}=\left( 
\begin{array}{cc}
0 & \sigma ^{i} \\ 
\sigma ^{i} & 0%
\end{array}
\right) ,
\end{equation*}
where $\sigma ^{i}$ are the Pauli matrices: 
\begin{eqnarray*}
\sigma ^{1} =\left( 
\begin{array}{cc}
0 & 1 \\ 
1 & 0%
\end{array}
\right) ,\qquad \sigma ^{2}=\left( 
\begin{array}{cc}
0 & -i \\ 
i & 0%
\end{array}
\right) ,\qquad \sigma ^{3} =\left( 
\begin{array}{cc}
1 & 0 \\ 
0 & -1%
\end{array}
\right).
\end{eqnarray*}

To investigate the modification of the energy levels by eq. \eqref{a2}, we
use the first-order perturbation theory, where, by restoring the constants $%
c $ and $\hbar $, the spectrum of $\hat{H}_{0}$ and the corresponding wave
functions are well-known and are given by (see \cite{34,39,40,41,42,43,44}): 
\begin{equation}
\psi \left( r,\theta ,\varphi \right) =\left( 
\begin{array}{c}
\phi \left( r,\theta ,\varphi \right) \\ 
\chi \left( r,\theta ,\varphi \right)%
\end{array}%
\right) =\left( 
\begin{array}{c}
f\left( r\right) \Omega _{jlM}\left( \theta ,\varphi \right) \\ 
g\left( r\right) \Omega _{jlM}\left( \theta ,\varphi \right)%
\end{array}%
\right) ,
\end{equation}%
where the bi-spinors $\Omega _{jlM}\left( \theta ,\varphi \right) $ are
defined by: 
\begin{equation}
\Omega _{jlM}\left( \theta ,\varphi \right) =\left( 
\begin{array}{c}
\mp \sqrt{\frac{\left( j+1/2\right) \mp \left( M-1/2\right) }{2j+\left( 1\pm
1\right) }}Y_{j\pm 1/2,M-1/2}\left( \theta ,\varphi \right) \\ 
\sqrt{\frac{\left( j+1/2\right) \pm \left( M+1/2\right) }{2j+\left( 1\pm
1\right) }}Y_{j\pm 1/2,M+1/2}\left( \theta ,\varphi \right)%
\end{array}%
\right) ,
\end{equation}%
with the radial functions $f\left( r\right) $ and $g\left( r\right) $ given
as: 
\begin{eqnarray}
\left( 
\begin{array}{c}
f\left( r\right) \\ 
g\left( r\right)%
\end{array}%
\right) &=&\left( a\frac{mc}{\hbar }\right) ^{2}\frac{1}{\nu }\sqrt{\frac{%
\hbar c\left( E\varkappa -m_{e}c^{2}\nu \right) n!}{\left( m_{e}c^{2}\right)
^{2}\alpha \left( \varkappa -\nu \right) \Gamma \left( n+2\nu \right) }}e^{-%
\frac{1}{2}x}x^{\nu -1}\times  \notag \\
&&\times \left( 
\begin{array}{c}
f_{1}xL_{n-1}^{2\nu +1}\left( x\right) +f_{2}L_{n}^{2\nu -1}\left( x\right)
\\ 
g_{1}xL_{n-1}^{2\nu +1}\left( x\right) +g_{2}L_{n}^{2\nu -1}\left( x\right)%
\end{array}%
\right) ,
\end{eqnarray}%
where the ordinary relativistic energy levels are given by: 
\begin{equation}
E=E_{n,j}=\frac{m_{e}c^{2}\left( n+\nu \right) }{\sqrt{\alpha ^{2}+\left(
n+\nu \right) ^{2}}},\qquad n=0,1,2\cdots
\end{equation}%
and $L_{n}^{\alpha }\left( x\right) $ are the associated Laguerre
polynomials \cite{31}, with the following notations: 
\begin{eqnarray*}
a=\frac{1}{m_{e}c^{2}}\sqrt{\left( m_{e}c^{2}\right) ^{2}-E^{2}}, &\qquad &x=%
\frac{2}{\hbar c}\sqrt{\left( m_{e}c^{2}\right) ^{2}-E^{2}}\,\,r, \\
\varkappa =\pm \left( j+\frac{1}{2}\right) , &\qquad &\nu =\sqrt{\varkappa
^{2}-\alpha ^{2}}, \\
f_{1}=\frac{a\alpha }{\frac{E}{m_{e}c^{2}}\varkappa -\nu }, &\qquad
&f_{2}=\varkappa -\nu , \\
g_{1}=\frac{a\left( \varkappa -\nu \right) }{\frac{E}{m_{e}c^{2}}\varkappa
-m_{e}\nu }, &\qquad &g_{2}=\frac{e^{2}}{\hbar c}=\alpha .
\end{eqnarray*}%
In the above $m_{e}$ is the mass of the electron and $\alpha $ is the fine
structure constant.

\subsection{Non-commutative corrections to the Dirac energy}

Now to obtain the modification to the energy levels as a result of the terms %
\eqref{a2} due to the non-commutativity of time-space, we use perturbation
theory up to the first order. With respect the selection rule $\Delta l=0$
and choosing the coordinate system $\left( t,r,\theta ,\varphi \right) $ so
that $\Theta ^{0k}=-\Theta ^{k0}=\Theta \delta ^{01}$, we have: 
\begin{equation}
\Delta E_{n,j}^{\left( \Theta \right) }=\Delta E_{n,j}^{\left( 1\right)
}+\Delta E_{n,j}^{\left( 2\right) },  \label{eq:energymod}
\end{equation}%
where: 
\begin{align}
\Delta E_{n,j}^{\left( 1\right) }& =\frac{E}{2\hbar c}e^{2}\int_{0}^{4\pi
}\Theta d\Omega \int_{0}^{\infty }dr[\psi _{njlM}^{\dagger }\left( r,\theta
,\varphi \right) \psi _{nj^{\prime }l^{\prime }M^{\prime }}\left( r,\theta
,\varphi \right) ]  \notag \\
& =\frac{E}{2\hbar c}e^{2}\Theta _{MM^{\prime }}\langle \frac{1}{r^{2}}%
\rangle ,
\end{align}
and 
\begin{align}
\Delta E_{n,j}^{\left( 2\right) }& =-\frac{e^{4}}{\hbar c}\int_{0}^{4\pi
}\Theta d\Omega \int_{0}^{\infty }drr^{-2}[\psi _{njlM}^{\dagger }\left(
r,\theta ,\varphi \right) \psi _{nj^{\prime }l^{\prime }M^{\prime }}\left(
r,\theta ,\varphi \right) ]  \notag \\
& =-\frac{e^{4}}{\hbar c}\Theta _{MM^{\prime }}\langle \frac{1}{r^{3}}%
\rangle ,
\end{align}
where: 
\begin{eqnarray}
\langle \frac{1}{r^{2}}\rangle &=&2\hbar c\left( \frac{m_{e}c}{\hbar }
a\right) ^{3}\left[ \frac{\varkappa \left( 2E\varkappa -m_{e}c^{2}\right) }{
\left( m_{e}c^{2}\right) ^{2}\alpha \nu \left( 4\nu ^{2}-1\right) }\right] ,
\\
\langle \frac{1}{r^{3}}\rangle &=&\left( \frac{mc}{\hbar }a\right) ^{3}\left[
\frac{3E\varkappa \left( E\varkappa -m_{e}c^{2}\right) }{\left(
m_{e}c^{2}\right) ^{2}\nu \left( 4\nu ^{2}-1\right) \left( \nu ^{2}-1\right) 
}\right.  \notag \\
&&\qquad \qquad\left. -\frac{\left( m_{e}c^{2}\right) ^{2}\left( \nu
^{2}-1\right) }{ \left( m_{e}c^{2}\right) ^{2}\nu \left( 4\nu ^{2}-1\right)
\left( \nu ^{2}-1\right) }\right] .
\end{eqnarray}

From equation \eqref{eq:energymod} we obtain the modified energy levels in
non-commutative space-time to the first order of $\Theta $ as: 
\begin{multline}
\Delta E_{n,j}^{\left( \Theta \right) }=\left( \frac{\alpha }{\hbar c}
\right) ^{2}\frac{m_{e}c^{2}a^{3}}{\nu \left( 4\nu ^{2}-1\right) }\times \\
\times \left[ E\varkappa \left( \frac{\left( 2E\varkappa -m_{e}c^{2}\right) 
}{\alpha ^{2}}-\frac{3\left( E\varkappa -m_{e}c^{2}\right) }{\left( \nu
^{2}-1\right) }\right) +\left( m_{e}c^{2}\right) ^{2}\right] \Theta
_{MM^{\prime }}.  \label{cc1}
\end{multline}

The selection rules for the transitions between the levels $\left(
Nl_{j}^{M}\rightarrow Nl_{j}^{M^{\prime }}\right) $\ are $\Delta l=1$\ and $%
\Delta M=0,\pm 1$, where $N=n+\left\vert \varkappa \right\vert $\ describes
the principal quantum number. The $2S_{1/2}$\ and $2P_{1/2}$\ levels
correspond respectively to: 
\begin{equation}
\left( N=1,j=1/2,\varkappa =\pm 1,M=\pm 1/2\right)  \label{aa1}
\end{equation}
From eqs. \eqref{cc1} and \eqref{aa1} we can write: 
\begin{eqnarray}
\Delta E_{2S_{1/2}}^{\left( \Theta \right) } &=& 1.944\,64\times
10^{-8}\Theta \,\left( \mathrm{MeV}\right) ^{3},  \label{tt} \\
\Delta E_{2P_{1/2}}^{\left( \Theta \right) } &=&\pm 2.160\,75\times
10^{-9}\Theta \,\left( \mathrm{MeV}\right) ^{3}.
\end{eqnarray}
The non-commutative correction to the transition follows as: 
\begin{equation*}
\Delta E_{2,1/2}^{\left( \Theta \right) }\left( 2P_{1/2}\rightarrow
2S_{1/2}\right) =2.160715\times 10^{-8}\Theta \text{ }\left( \mathrm{MeV}%
\right) ^{3}
\end{equation*}

Now again using the current theoretical result on the $2P$ Lamb shift from
ref. \cite{38} which is about $0.08$ kHz, and from the splitting \eqref{tt},
we get the bound: 
\begin{equation}
\Theta \lesssim \left( 0.25\,\mathrm{TeV}\right) ^{-2}.  \label{eq:bound}
\end{equation}%
Restoring the constants $c$ and $\hbar $ in eq. \eqref{eq:bound} we write
the bound on the non-commutativity parameter as: 
\begin{equation}
\Theta \lesssim 1.7\times 10^{-35}\mathrm{m}^{2}.
\end{equation}

It is interesting that the value of the upper bound on the time-space
non-commutativity parameter as derived here is better than the results of
refs. \cite{12, 21, 35, 45} . This value is only in the sense of an upper
bound and not the value of the parameter itself, for which the fundamental
length $\sqrt{\Theta }$ is compatible with the value of the electroweak
length scale $\left( \ell _{\omega }\right) $. This effectively confirms the
presence of gravity at this level.

\section{Conclusions}

In this work we started from quantum relativistic charged scalar and fermion
particles in a canonical non-commutative space-time to find the action which
is invariant under the infinitesimal gauge transformations. By using the
Seiberg-Witten maps and the Moyal product , we have derived the deformed KG
and Dirac equations for non-commutative Coulomb potential up to first order
in the non-commutativity parameter $\Theta $. By solving the deformed KG and
Dirac equations we found the $\Theta $-correction energy shift. This proves
that the non-commutativity has an effect similar to that of the magnetic
field. The corrections induced to the energy levels by this non-commutative
effect and the lamb-shift were induced and compared to experimental results
from high precision hydrogen spectroscopy to obtain a new bound for the
non-commutativity parameter of around $\left( 0.25\,\mathrm{TeV}\right)
^{-2} $, for which the fundamental length $\sqrt{\Theta }$ is compatible
with the value of the electroweak length scale $\left(\ell _{\omega }\right)$%
. This effectively confirms the presence of gravity at this level.

\end{document}